\documentclass[intlimits,twoside,a4paper]{article}
\usepackage{amsmath,amssymb}
\usepackage{graphicx}
\usepackage{wrapfig}
\usepackage[T2A]{fontenc}
\usepackage[cp1251]{inputenc}
\usepackage{cmpj}

\issue{2011}{14}{3}{33005}

\doinumber{10.5488/CMP.14.33005}

\def\be{\begin{equation}}
\def\ee{\end{equation}}
\def\bea{\begin{eqnarray}}
\def\eea{\end{eqnarray}}

\def\gsim{\mathrel{\lower.65ex\hbox{$\mathop{\kern0pt\sim}\limits
   ^{\lower.55ex\hbox{$>$}}$}}}
\def\lsim{\mathrel{\lower.65ex\hbox{$\mathop{\kern0pt\sim}\limits
   ^{\lower.55ex\hbox{$<$}}$}}}

\title[MSA for the Lennard-Jones-like two Yukawa model]
{Mean-spherical approximation for the Lennard-Jones-like two Yukawa model:
Comparison against Monte Carlo data
}

\author[I. Nezbeda \textsl{et al.}]{J.~Krej\v{c}\'{i}\refaddr{IOP},
I.~Nezbeda\refaddr{IOP,LCI}, R.~Melnyk\refaddr{IP}, A.~Trokhymchuk\refaddr{IP,BYU}
}

\authorcopyright{J. Krej\v{c}\'{i},  I. Nezbeda, R. Melnyk, A. Trokhymchuk, 2011}

\addresses{
\addr{IOP} Faculty of Science, J. E. Purkinje University, 400~96 \'{U}st\'\i \ n. Lab., Czech Republic
\addr{LCI} E. H\'{a}la Laboratory of Thermodynamics, ICPF, Acad.\ Sci., 165~02~Prague~6~--~Suchdol, Czech Republic
\addr{IP} Institute for Condensed Matter Physics
of the National Academy of Sciences of Ukraine,\\
1 Svientsitskii Str., 79011 Lviv, Ukraine
\addr{BYU} Department of Chemistry and Biochemistry, Brigham Young University, Provo, UT 84602, USA
}
\date{Received August 1, 2011}%

\begin{document}

\maketitle

\begin{abstract}
Monte Carlo simulation studies are performed for
the Lennard-Jones-like two Yukawa (LJ2Y) potential to show how properties of
this model fluid depend on the replacement of soft repulsion by hard-core
repulsion. Different distances for the positioning of hard-core have been
explored. We have found that for temperatures slightly lower and
slightly higher than the critical point temperature for the Lennard-Jones fluid, the placement of
the hard-core at distances shorter than zero-potential energy is
well justified by thermodynamic properties that are practically the same as in
the original LJ2Y model without hard-core. However, going to extreme conditions with
the high temperature one should be careful since the presence of the hard-core provokes
changes in the properties of the system. The later is extremely important when the
mean-spherical approximation (MSA) theory is applied to the treatment of the Lennard-Jones-like
fluid.
\keywords two Yukawa potential, Lennard-Jones fluid, mean-spherical approximation,
Monte Carlo simulations
\pacs 01.65.Q
\end{abstract}

\section{Introduction}

The so-called simple fluid models cannot be always applied to the study of realistic systems.
However, this type of models is very important in basic science while studying
fundamental problems in the liquid state theory. One of these problems refers to the
role played by repulsive and attractive forces. While comprehension of the repulsive
forces is due to the progress reached in the theoretical and computer modeling of
 hard sphere fluids, the simplest nontrivial model that makes it possible to study the liquid
phase of a matter as well as the vapor-liquid coexistence seems to be the one that
consists of a hard-core plus attractive Yukawa (HCAY) potential
\begin{eqnarray}
u_{\rm 1Y}(r)/\epsilon
 &=& \infty \hspace{31mm} {\rm for}\,\,  r < R \nonumber\\
 &=& -\frac{R}{r}\exp [\kappa (r -R)] \qquad
     {\rm for} \,\, r \geqslant  R,
\label{uY}
\end{eqnarray}
where $\epsilon$ is the depth of the potential energy well, $R$ is a hard-core diameter,
$\kappa^{-1}$ is a measure of the range of the attractive tail.

\begin{figure}[ht]
\begin{center}
\includegraphics[width=7.00cm]{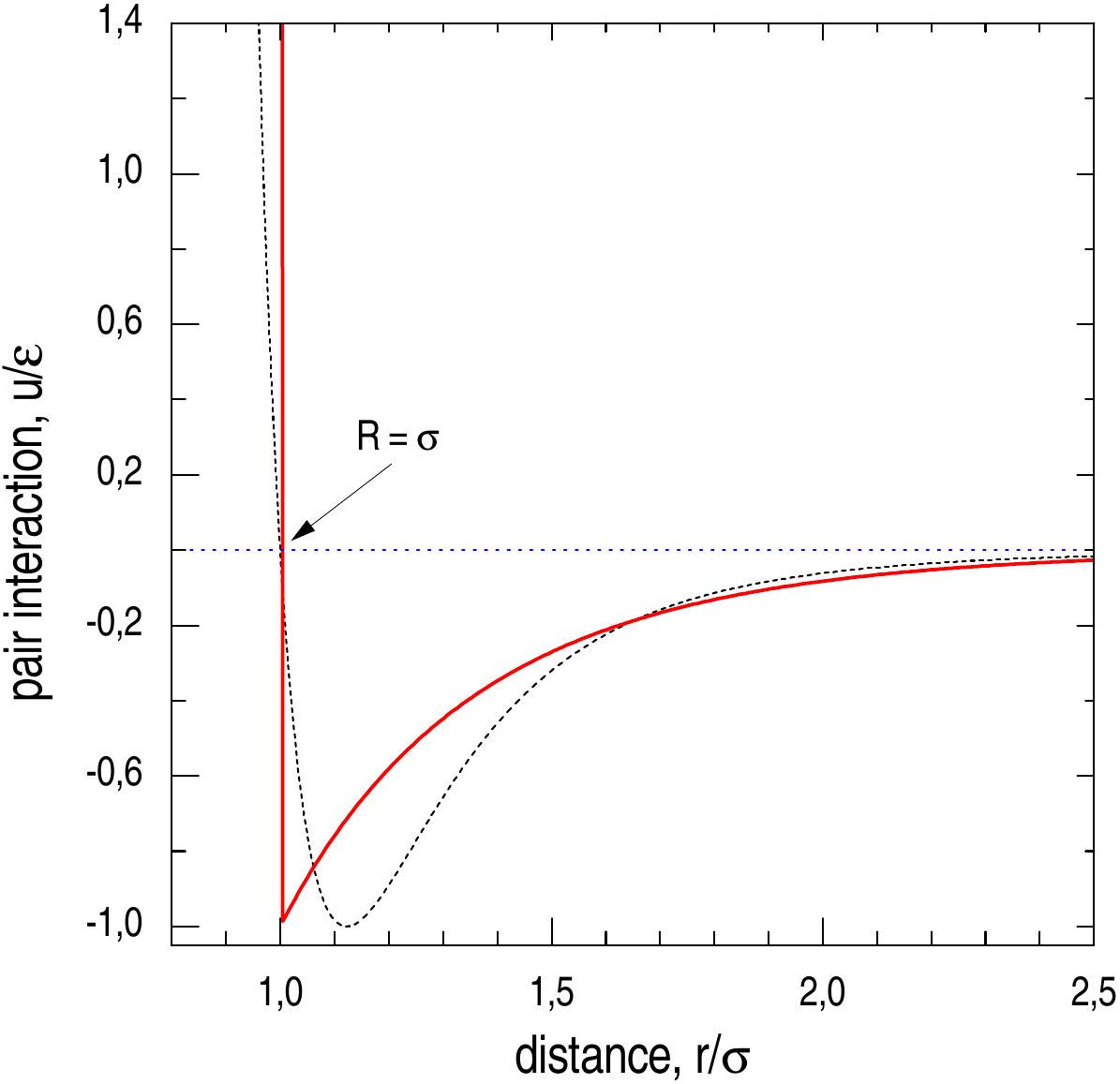}
\hspace{0.5cm}
\includegraphics[width=7.00cm]{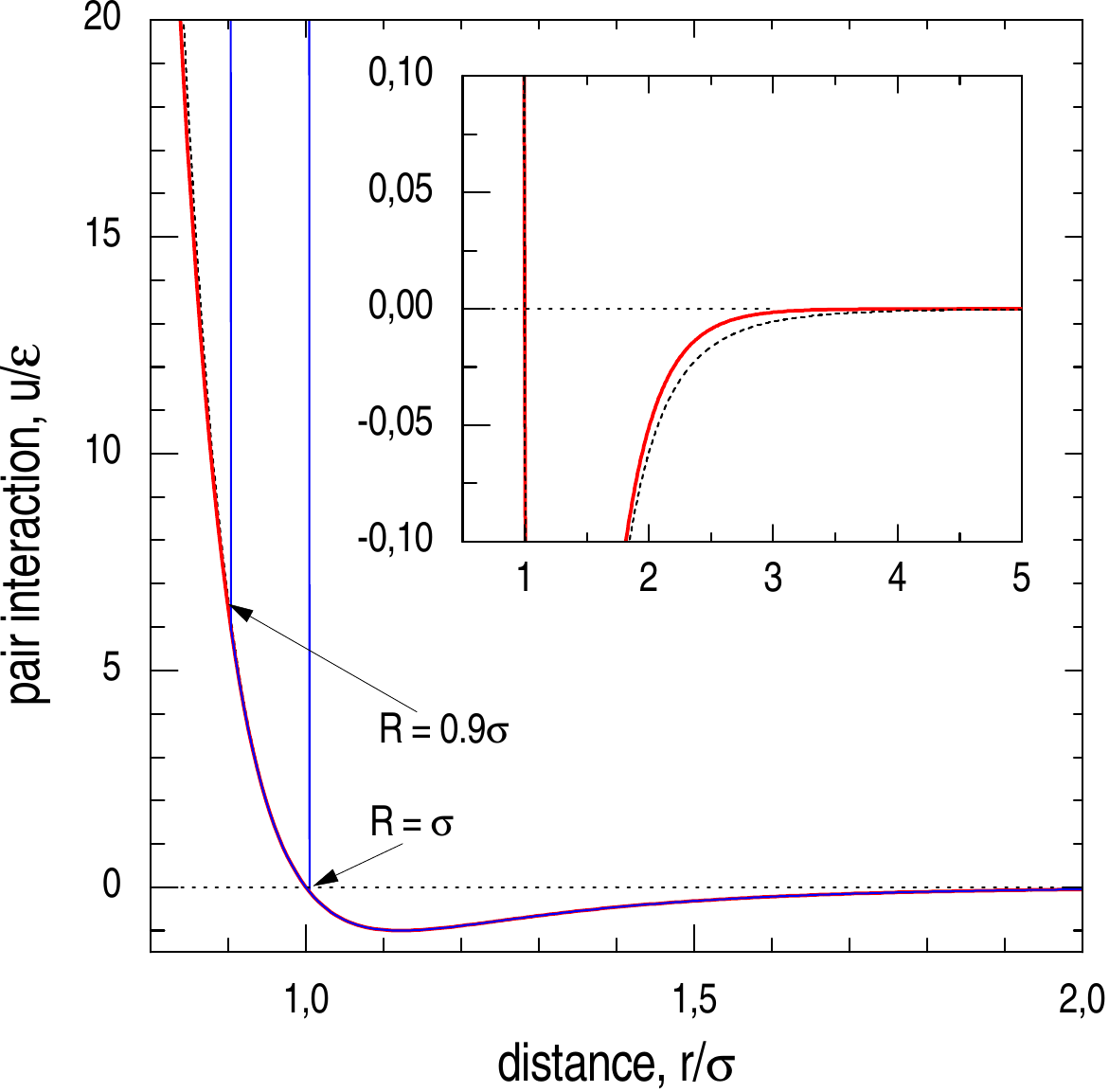}
\end{center}
\caption{The Lennard-Jones potential in comparison with the
hard-core attractive Yukawa (HCAY) potential (left part) and with
the Lennard-Jones-like two Yukawa (LJ2Y) potential with different
positions of the hard-core (right part).}
\label{fig1}
\end{figure}

An important feature due to which the HCAY model is intensively employed in the studies
of simple fluids
is that approximate analytical and semianalytical solutions for this model are
available owing to the pioneering paper by Waisman  \cite{waisman} on the analytic solution
of the Ornstein-Zernike equation using the mean-spherical approximation (MSA).
These were Henderson et al.~\cite{henderson}
who by using this MSA solution suggested that having $R=\sigma$ and
$\kappa R=1.8$, the HCAY fluid is qualitatively
similar to argon for the densities and temperatures of the liquid in equilibrium
with its vapour and with  the  potential parameters approximately the same as the Lennard-Jones potential (i.e. $\epsilon/k ~ 120~K, \sigma\approx 3.4$~\AA).
Figure~\ref{fig1}~(a) shows this Lennard-Jones-like HCAY fluid in comparison with the original
Lennard-Jones (LJ) fluid.

Recently, Kadiri et al.~\cite{Kadiri2008} have extended the mapping of the LJ system
into the HCAY system over a wide domain of a phase diagram.
Specifically, at every density and temperature the thermodynamics of the LJ fluid
was reproduced from the equation of state of the HCAY fluid with suitable values
of its two parameters $R$ and $\kappa$.
An advantage of such an approach is that it provides analytical equations
for the thermodynamics of the LJ system with only two parameters
that are of precision comparable to that
of the simulation data. It can be compared to the fully empirical equations
of state by Johnson et al.~\cite{Jonson93} that uses 33 parameters.
However, one can see from figure~\ref{fig1} that the shapes of two potential profiles, LJ and 1Y,
are rather different and one cannot be sure that the thermodynamics predicted
by two models will be always the same.
Thus, this approach should be used with great care while
exploring the thermodynamic states
outside of the mapping domain.

More possibilities are offered by the MSA solution that has been obtained by
Blum and Hoye~\cite{BlumHoye} for a linear superposition of the Yukawa
tails $\sum\epsilon_{i}\exp(-\kappa_{ i}r)/r$.
This was quite an important step since superposition of the
attractive and repulsive Yukawa tails makes it possible to
mimic  practically any potential profile. It is natural that this MSA solution
has been applied to  Lennard-Jones fluid represented by the hard core and  sum of two
Yukawas (HC2Y)~\cite{Konior1988,Yukal1996},
\begin{eqnarray}
u_{\rm 2Y}(r)/\epsilon
 &=& \infty \hspace{66mm} {\rm for}\,\,  r < R \nonumber\\
 &=& \frac{R}{r} \exp [-\kappa_1(r-R)] -
  \frac{R}{r} \exp [-\kappa_2(r-R)] \qquad {\rm for} \,\, r \geqslant  R,
\label{u2Ymsa}
\end{eqnarray}
where again $R=\sigma$ determines the separation of zero energy, $u(\sigma)=0$,
and $\epsilon$ is, as usual, the depth of the potential minimum.
The detailed course of the 2Y potential
(with a specific set of the decay parameters $\kappa_{\rm i}$ which will be
discussed in the following section~2)
is shown in figure~\ref{fig1}~(b) where it is also compared with the
Lennard-Jones (LJ) potential. Indeed, two Yukawa terms allow us to significantly
improve the mimicking of LJ potential at short and intermediate distances.
This is a rather important achievement since in this case the differences in the properties of these
two model systems can be attributed to the long-ranged asymptote of the LJ interaction only.
Another important issue is that in general, the LJ2Y potential (like the parent LJ potential)
does not require the presence of hard-core, e.g., for computer simulation studies;
the hard-core is necessary if one intends
to apply the MSA theory to the treatment of the problem.
However, even in this case placing the hard-core at $R=\sigma$ (as it usually is assumed) seems not to be always
appropriate and justified since it may modify the properties of the original model.
To clarify this issue, we will report here the
Monte Carlo simulation data for the original LJ2Y potential without hard-core as well as for the
LJ2Y potential with a few different choices of the hard-core diameter $R$ in the range $0.8<R/\sigma<1$.
One more question that we are seeking to answer in this study concerns the performance
of MSA theory for different choices of the hard-core diameters $0.8<R/\sigma<1$.
The MSA studies reported so far for the LJ2Y model are dealing mainly with the case $R=\sigma$.

The structure of this paper is as follows.
In  section~2 we outline the way the parameters of the LJ2Y potential have been
chosen and present a brief description of the Monte Carlo computer simulations and
MSA computations that we have carried out for the LJ2Y model. The results are
collected and discussed in section~3 while section~4 contains conclusions.

\section{The potential models and computational details}

Originally, the LJ2Y potential is a model made up of two Yukawa tails without any hard-core,
\begin{equation}
u_{2Y}(r)/\epsilon =
 \epsilon_1\frac{r_{\rm m}}{r} \exp [-\kappa_1r] -
  \epsilon_2\frac{r_{\rm m}}{r} \exp [-\kappa_2r]\,,
\label{u2Y}
\end{equation}
where $r_{\rm m} $ is the position of the potential energy well,  $\epsilon_1>0$ and $\epsilon_2>0$ are the strengths of the
repulsive and attractive contributions, respectively, while $\kappa_1^{-1}$
and $\kappa_2^{-1}$ are the measures of the range of the corresponding
tails.
Superposition of two Yukawa tails in the form of equation~(\ref{u2Y})
to mimic the LJ potential has been
already used by other authors  \cite{tangfpe97,cummings,foilet}. However, all previous studies
have been concerned with the medium and long distances without paying much attention
to short separations governed by repulsive forces.

To map the properties of the LJ fluid onto those of a LJ2Y fluid, the above
four parameters of potential function (\ref{u2Y}) must be determined. Focussing
first on the course of the LJ function itself, the following three conditions
seem evident: the coincidence of the location and depth of the potential
minimum of the LJ and LJ2Y functions,
\begin{equation}
u_{\rm 2Y}(r=r_{\rm min}) = u_{\rm LJ}(r=r_{\rm min}) = -\epsilon\,,
\label{condition2}
\end{equation}
\begin{equation}
\frac{{\rm d}u_{\rm 2Y}(r)}{{\rm d}r}|_{r=r_{\rm min}}
 = \frac{{\rm d}u_{\rm LJ}(r)}{{\rm d}r}|_{r=r_{\rm min}} = 0\,\,,
\label{condition3}
\end{equation}
and the location of the potential zero,
\begin{equation}
u_{\rm 2Y}(r=\sigma)  =  u_{\rm LJ}(r=\sigma) = 0 \,\,.
\label{condition1}
\end{equation}
Then there remains to impose one more condition on the 2Y potential parameters
to complete the set of equations. Some authors attempt to follow the (12--6) LJ curve
to certain intermediate distance $r^*$ by equalizing the integrals from two potential functions
in the interval from $r_{\rm min}$ to $r^*$.
By contrast, in this study we will set the condition
that both potentials attain the same value at the location of the inflection point
$r=r_{\rm inf}=1.244455\sigma$ that follows immediately after the position of the potential well
of the LJ potential, i.e.,
\begin{equation}
u_{\rm 2Y}(r=r_{\rm inf}) = u_{\rm LJ}(r=r_{\rm inf})         \,\,.
\end{equation}

Proceeding in this way we obtain a 2Y potential that very accurately reproduces the LJ
potential profile at distances up to inflection point, but showing some discrepancies after
inflection point due to an exponential decay at large distances.
The resulting parameters are as follows: $\epsilon_1=1954325.046\sigma$,
$\epsilon_2=50.26984765\sigma$, $\kappa_1\sigma=13.66462$
and $\kappa_2\sigma=3.10147$. These values are consistent with those
in equation~(\ref{u2Ymsa}) and in figure~\ref{fig1}~(b).

\begin{figure}[ht]
\begin{center}
\includegraphics[width=7cm]{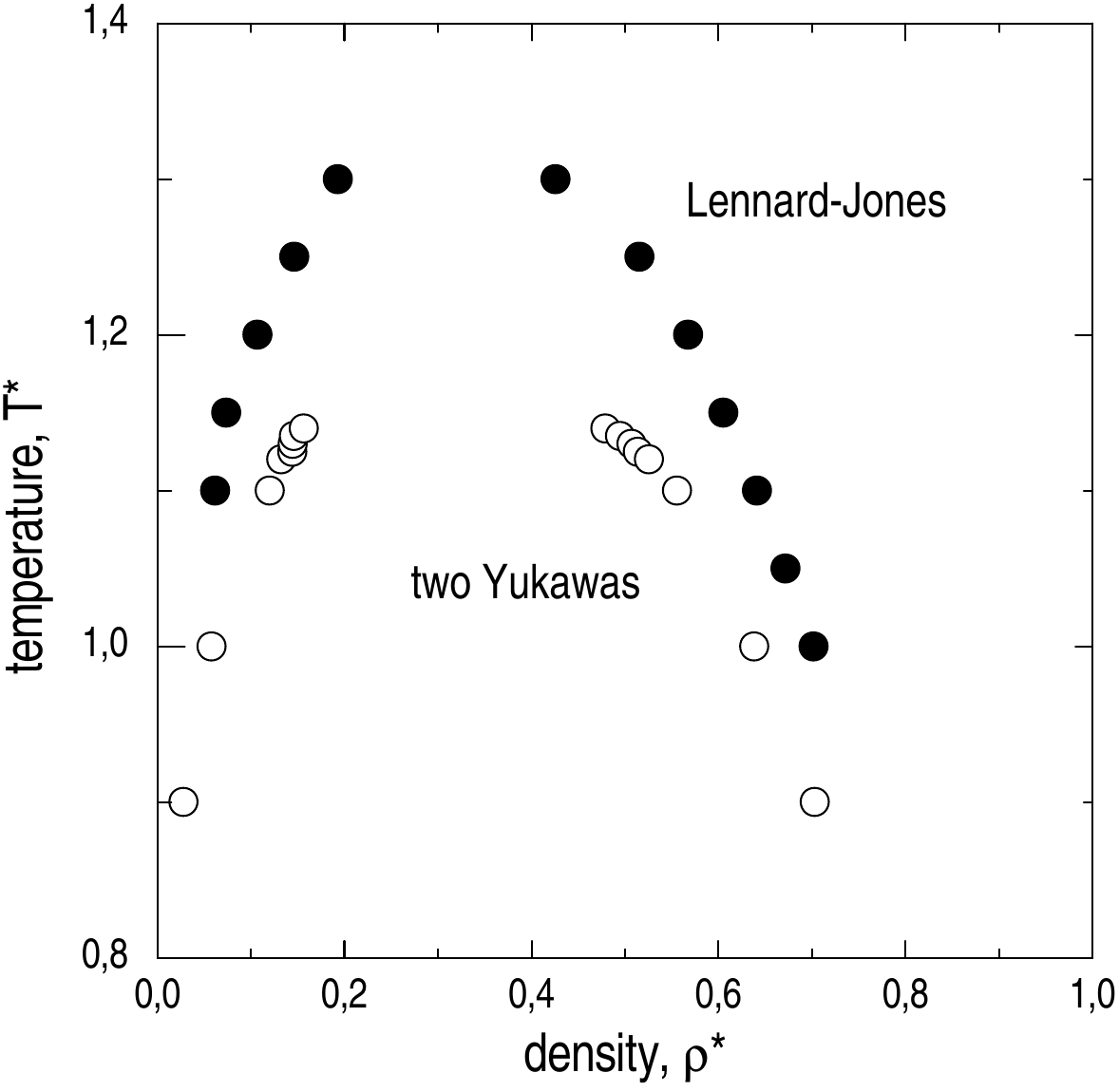}
\end{center}
\caption{The vapour-liquid coexistence envelope for the Lennard-Jones
fluid  \cite{lotfi} and for the Lennard-Jones-like two Yukawa (LJ2Y) fluid.}
\label{fig2}
\end{figure}
Figure~\ref{fig2} shows the vapour-liquid phase diagram for the LJ2Y potential model.
To determine the vapour-liquid envelope we used the
common Gibbs ensemble with the total number of particles $N = 512$ and applied the long-range
correction to truncate the potential at $r_{\rm c} = \sqrt[3]{N/\rho^*}$.
For the purpose of comparison we are presenting computer simulation data for the liquid-vapor
coexistence in a parent LJ fluid. The differences  that are observed in figure~\ref{fig2} concern
the lowering of the critical point temperature and
are caused by the differences between two potential models, LJ and LJ2Y, at
distances $r$ larger than the position of inflection point $r_{\rm inf}$.
These differences are consistent with what has been already learned from the computer simulation
studies of the HCAY model, namely, the shortening of the range of attraction brings about a
decrease of the critical point temperature. The corresponding MC data are collected in table~1.
\begin{table}[h]
\label{tab1}
{\small
\begin{center}
\caption{Monte Carlo data for vapour-liquid coexistence in the Lennard-Jones-like 2Y (LJ2Y) fluid.}
\vspace{0.25cm}
\begin{tabular}{r r r}
\hline
\multicolumn{1}{c}{$T^*$}&\multicolumn{1}{c}{$\rho^*_{\rm l}$}&\multicolumn{1}{c}{$\rho^*_{\rm v}$}\\
\hline
 0.900 & 0.0278$\pm$0.0084 & 0.7031$\pm$0.0133\\
 1.000 & 0.0579$\pm$0.0145 & 0.6385$\pm$0.0157\\
 1.100 & 0.1204$\pm$0.0278 & 0.5560$\pm$0.0169\\
 1.120 & 0.1324$\pm$0.0364 & 0.5259$\pm$0.0150\\
 1.125 & 0.1442$\pm$0.0373 & 0.5138$\pm$0.0169\\
 1.130 & 0.1448$\pm$0.0364 & 0.5074$\pm$0.0247\\
 1.135 & 0.1459$\pm$0.0493 & 0.4949$\pm$0.0267\\
 1.140 & 0.1568$\pm$0.0609 & 0.4792$\pm$0.0228\\
\hline
\end{tabular}
\end{center}}
\end{table}

To discuss the role played by repulsive forces of the hard-core origin we employed
both Monte Carlo (MC) computer simulations and MSA approaches.
First, we performed MC computer simulations for the original LJ2Y model given by equation~(\ref{u2Y})
for a set temperature and density conditions that are representative for LJ fluid model.
Then similar MC computer simulations at the same thermodynamic states have been repeated
for the LJ2Y model with an artificially embedded hard-core at different interparticle separations
smaller than LJ diameter $\sigma$.

In all these studies, the standard Monte Carlo simulations
for the LJ2Y model fluids were carried out in an NVT
ensemble with $N = 512$ particles using the largest possible cutoff for each
isotherm and with appropriate long range corrections applied~\cite{Frenkel}. In
addition to the common thermodynamic properties, the internal energy and
pressure (evaluated by the virtual volume change), we also determined, for all
thermodynamic state conditions considered, the excess chemical potential using
the standard Widom's particle insertion method~\cite{Frenkel}.
Our MSA study is based on the general solution obtained by Blum and Hoye~\cite{BlumHoye}
for a mixture of hard spheres
interaction via pair potential that is the sum of Yukawa tails with different decay
parameters. The numerical solution of the resulting set of nonlinear algebraic equation
was obtained using a relatively simple and effective iterative procedure proposed recently
 by Kalyuzhnyi and Cummings~\cite{KalCummings2006}.

Firstly, we are expecting that the placement of a hard-core at different distances may modify the
properties set up for the initial LJ2Y potential function given by equation~(\ref{u2Y}).
This issue can be resolved by analyzing the MC simulation data.
Secondly, the MSA performance could depend on the position of the hard-core, and
this can be concluded by comparing with computer simulation data.

\section{Results and discussions}

\begin{figure}[ht]
\vspace{4ex}
\begin{center}
\includegraphics[width=5.5cm]{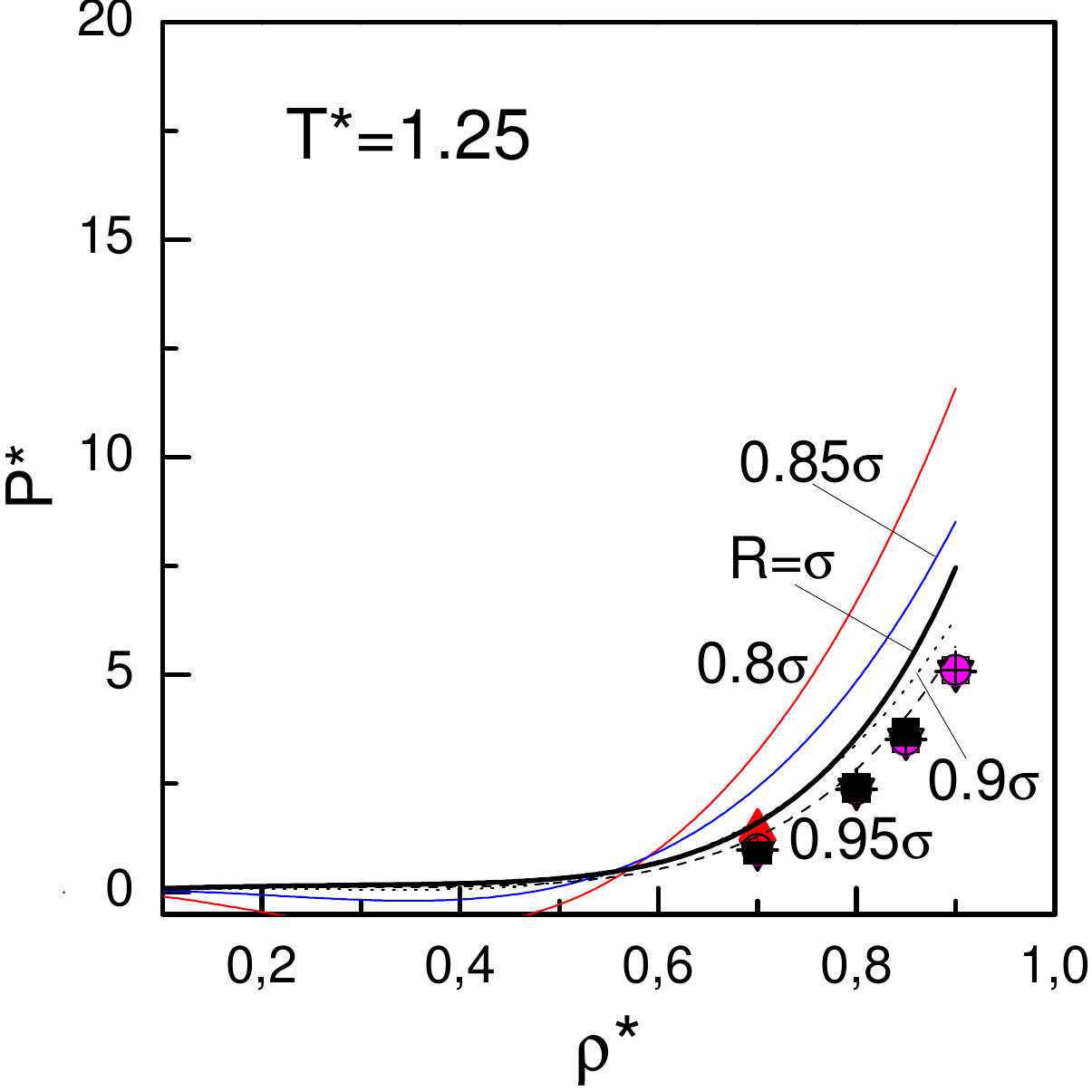}
\includegraphics[width=5.5cm]{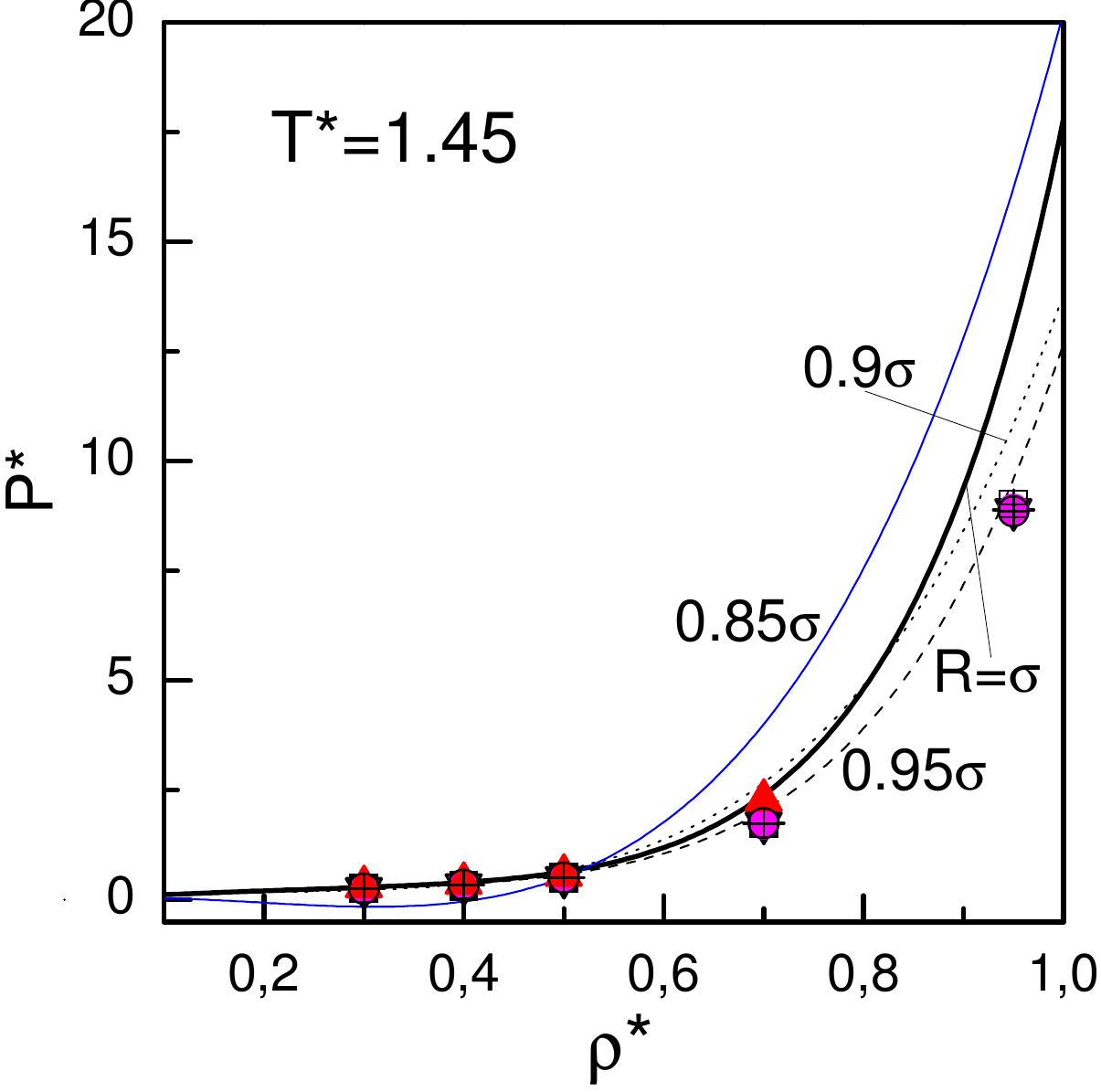}
\includegraphics[width=5.5cm]{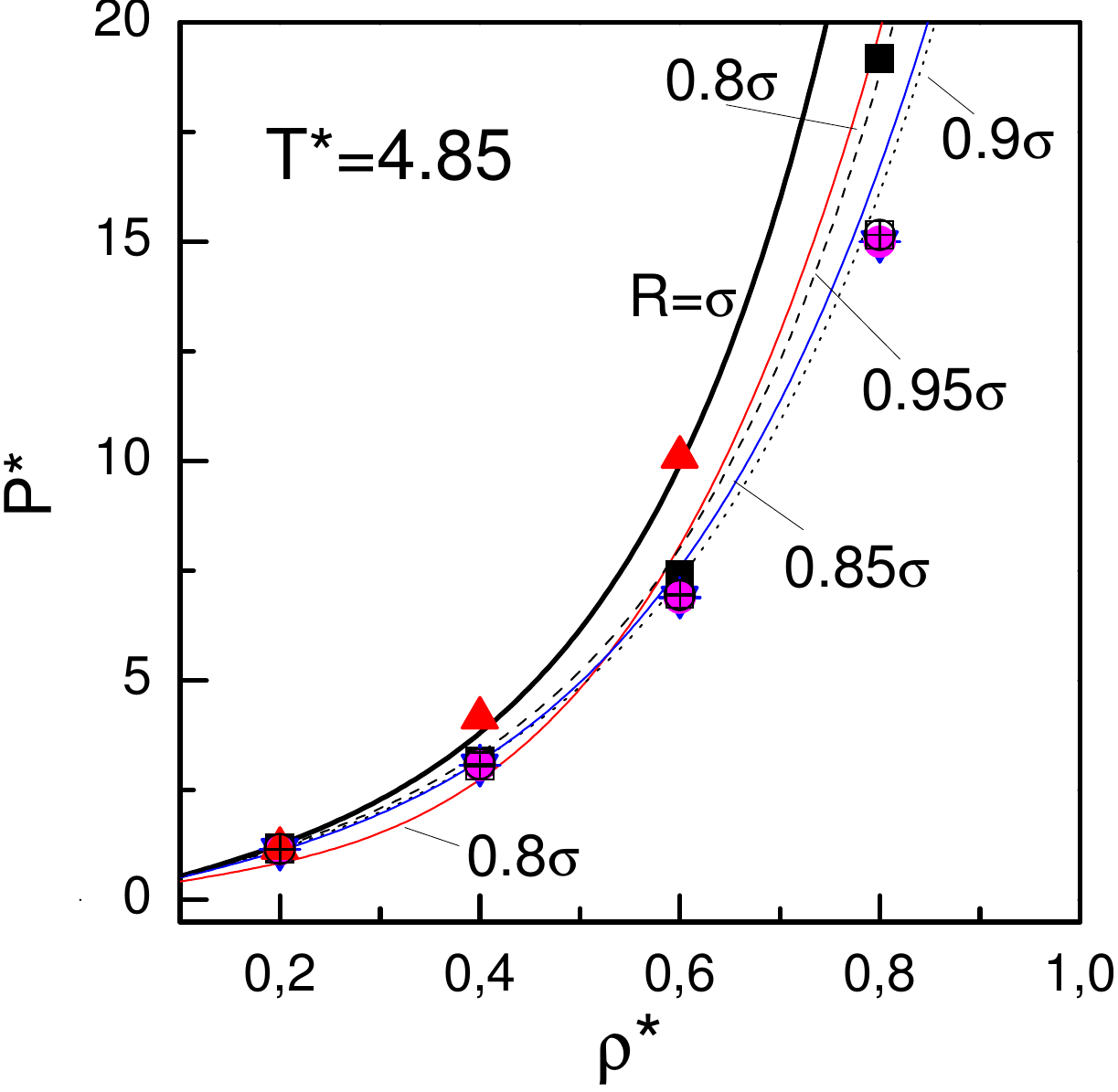}
\end{center}
\caption{The pressure of the LJ2Y fluid obtained from Monte Carlo computer simulations
(symbols) and in compliance with the  MSA theory (lines). The meaning of the symbols: filled triangles -- $R=\sigma$,
filled squares -- $R=0.95\sigma$,  crossed squares -- $R=0.9\sigma$, crossed circles -- $R=0.85\sigma$,
filled circles -- $R=0.8\sigma$, crossed triangles -- without hard-core.
The notes in the figures indicate the position of hard-core in MSA calculations.}
\label{fig3}
\end{figure}
\begin{figure}[ht]
\begin{center}
\includegraphics[width=5.5cm]{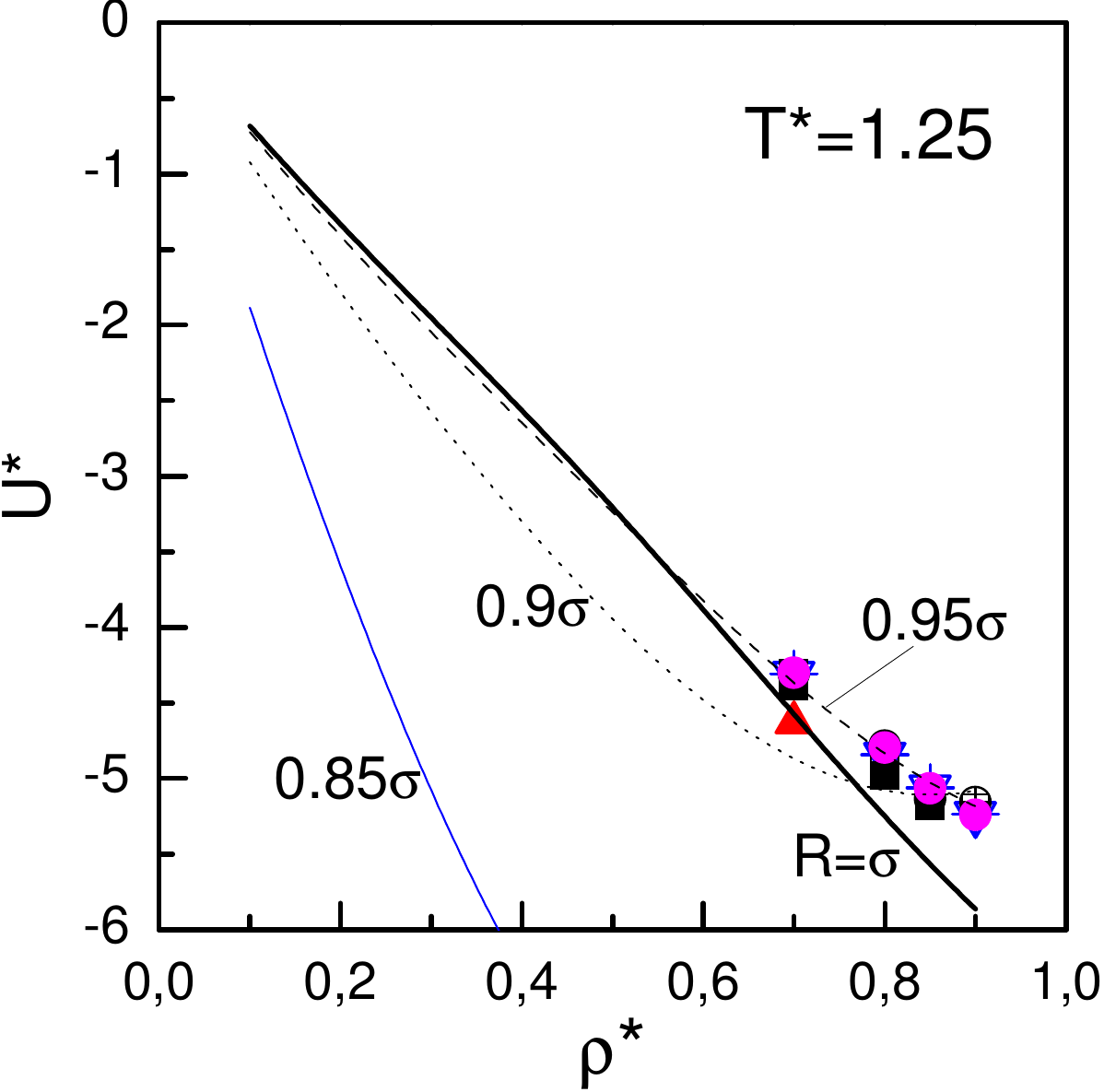}
\includegraphics[width=5.5cm]{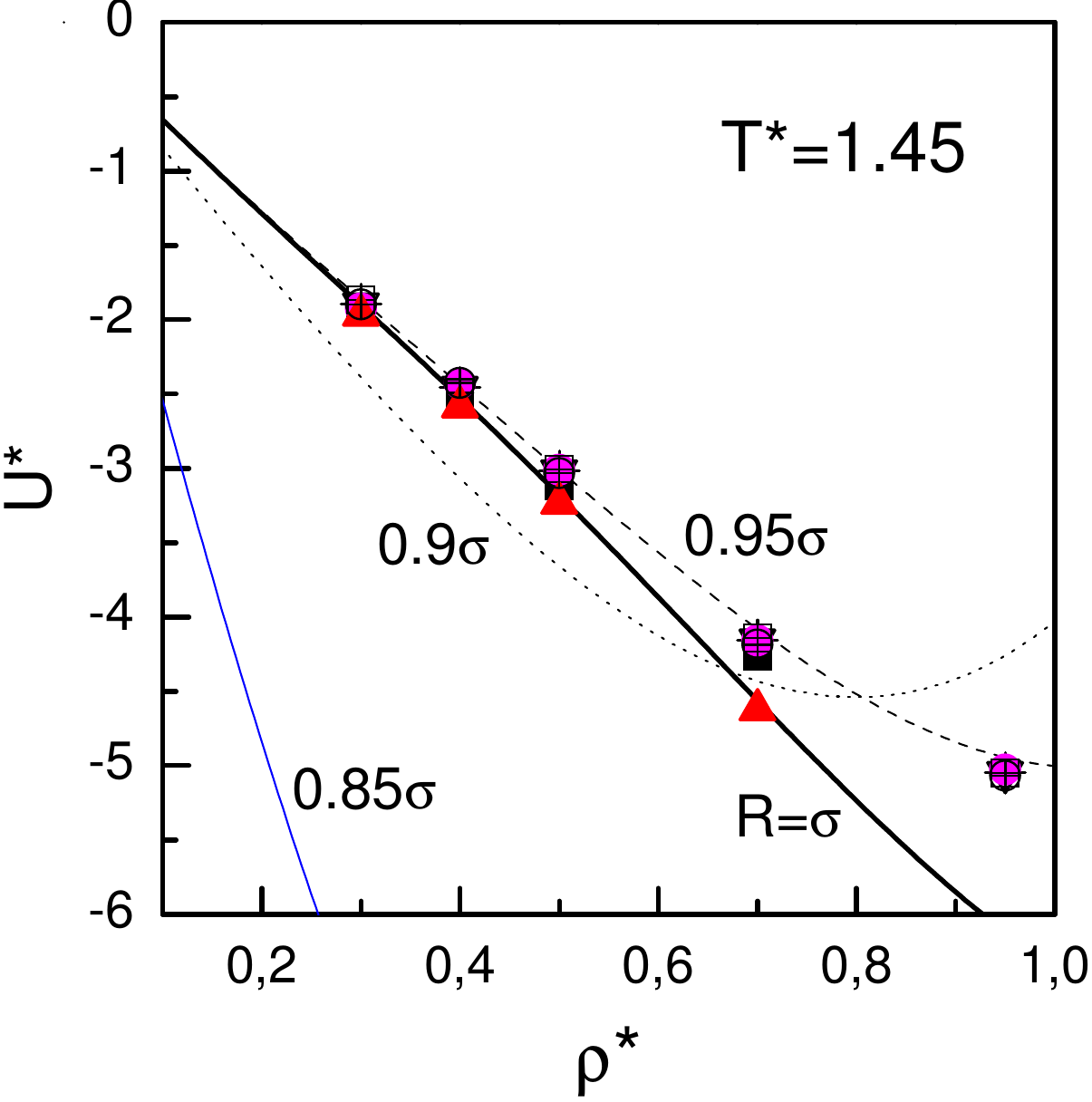}
\includegraphics[width=5.5cm]{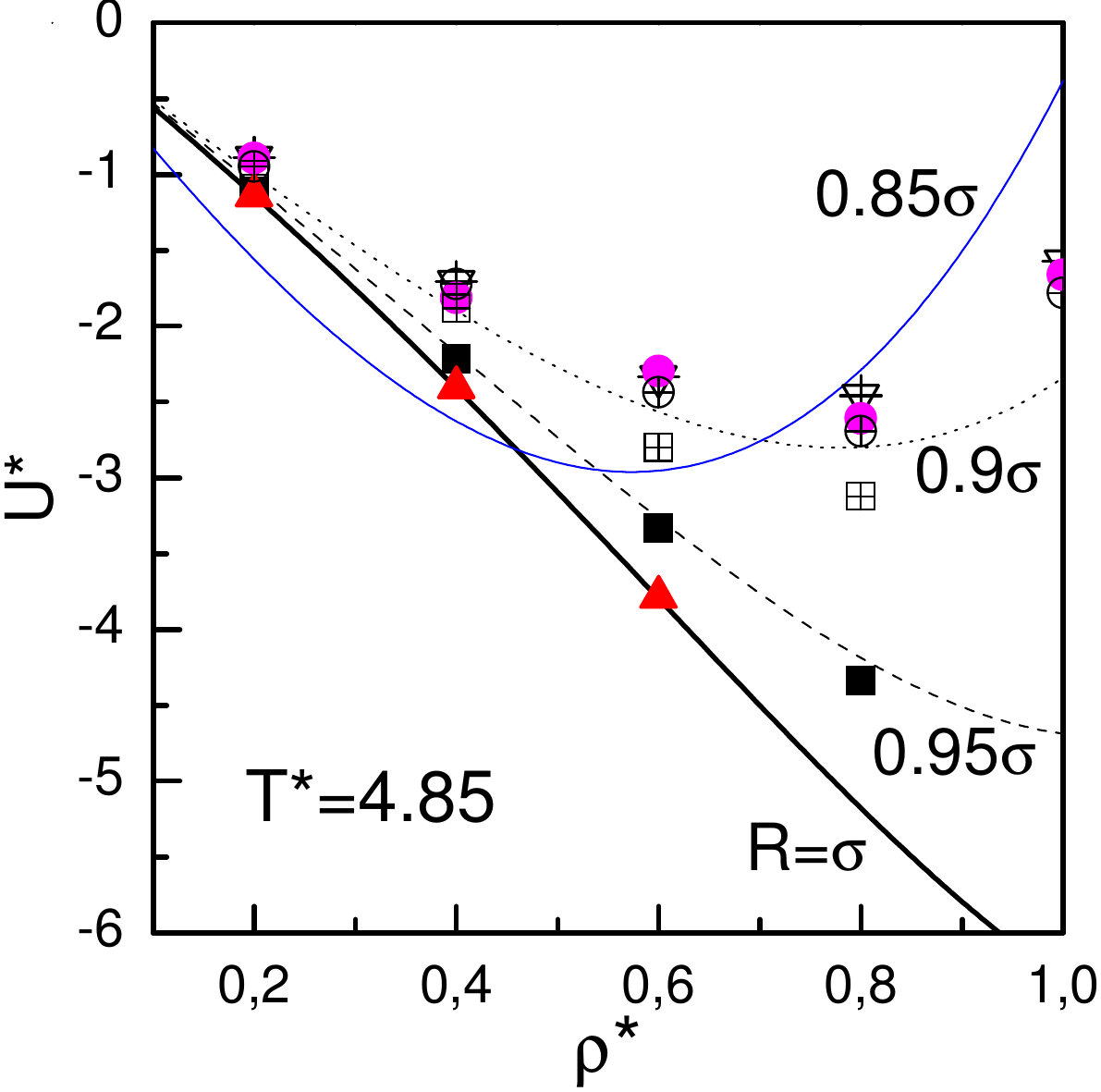}
\end{center}
\caption{The internal energy of the LJ2Y fluid obtained from Monte Carlo computer simulations
(symbols) and in compliance with the  MSA theory (lines). The meaning of the symbols is the same as in figure~\ref{fig3}.}
\label{fig4}
\end{figure}
\begin{figure}[!h]
\vspace{2ex}
\begin{center}
\includegraphics[width=7cm]{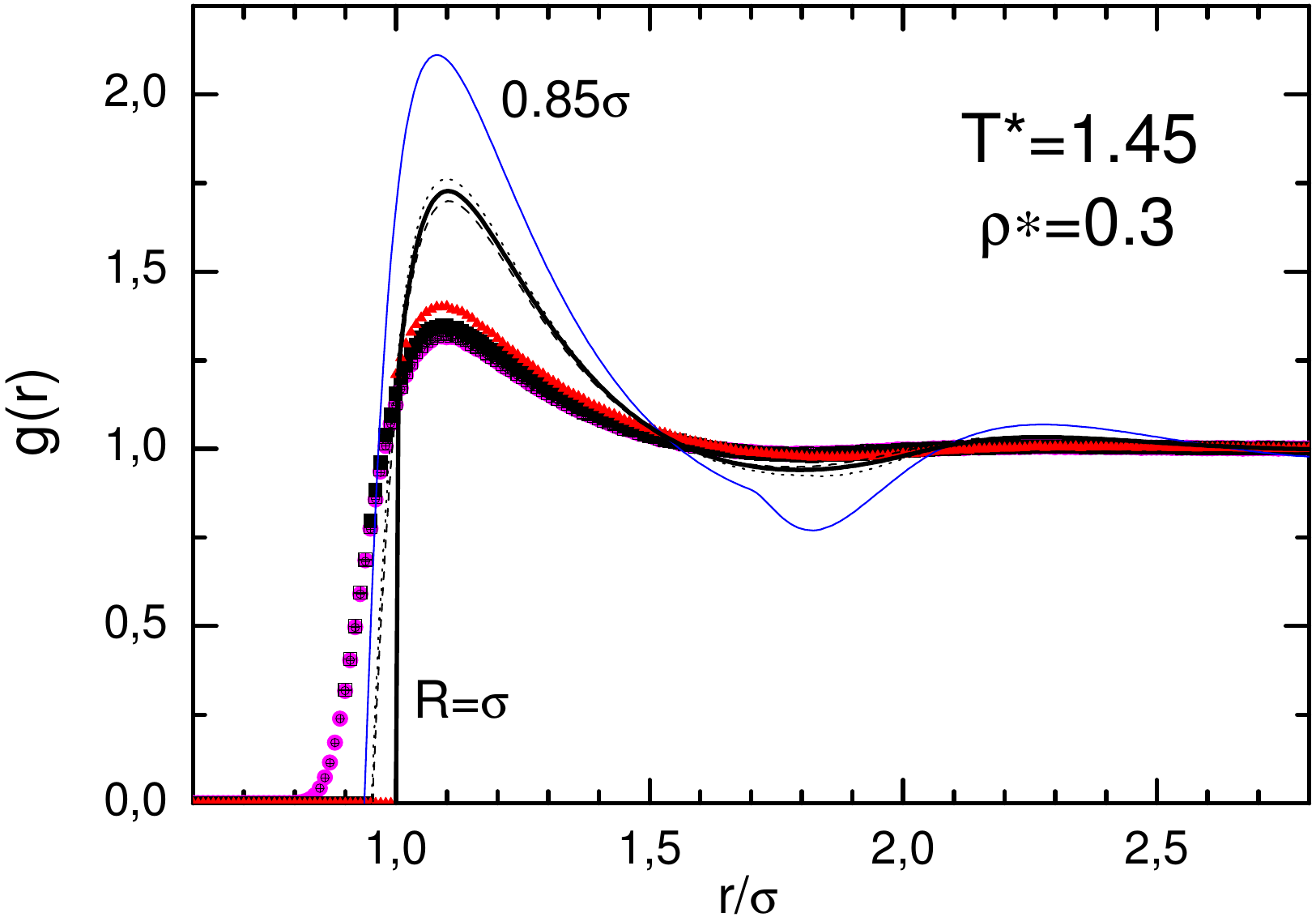}
\includegraphics[width=7cm]{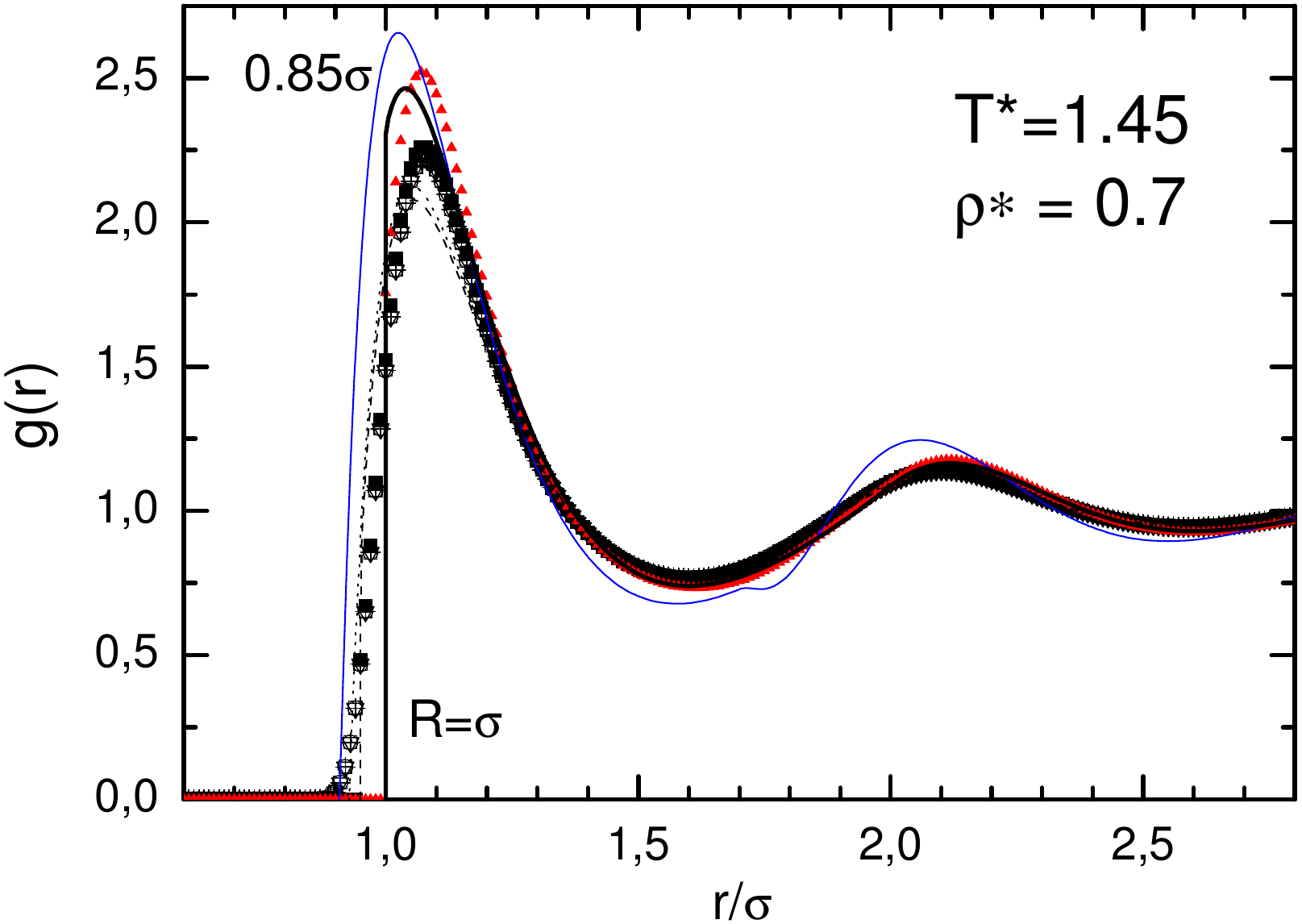}
\end{center}
\caption{The radial distribution functions of the LJ2Y fluid obtained from Monte Carlo computer simulations
(symbols) and in compliance with the  MSA theory (lines) at a slightly subcritical
temperature $T^*=1.45$ and two densities specified in the figure.
The meaning of the symbols is the same as in figure~\ref{fig3}.}
\label{fig5}
\end{figure}
\begin{figure}[!h]
\begin{center}
\includegraphics[width=7cm]{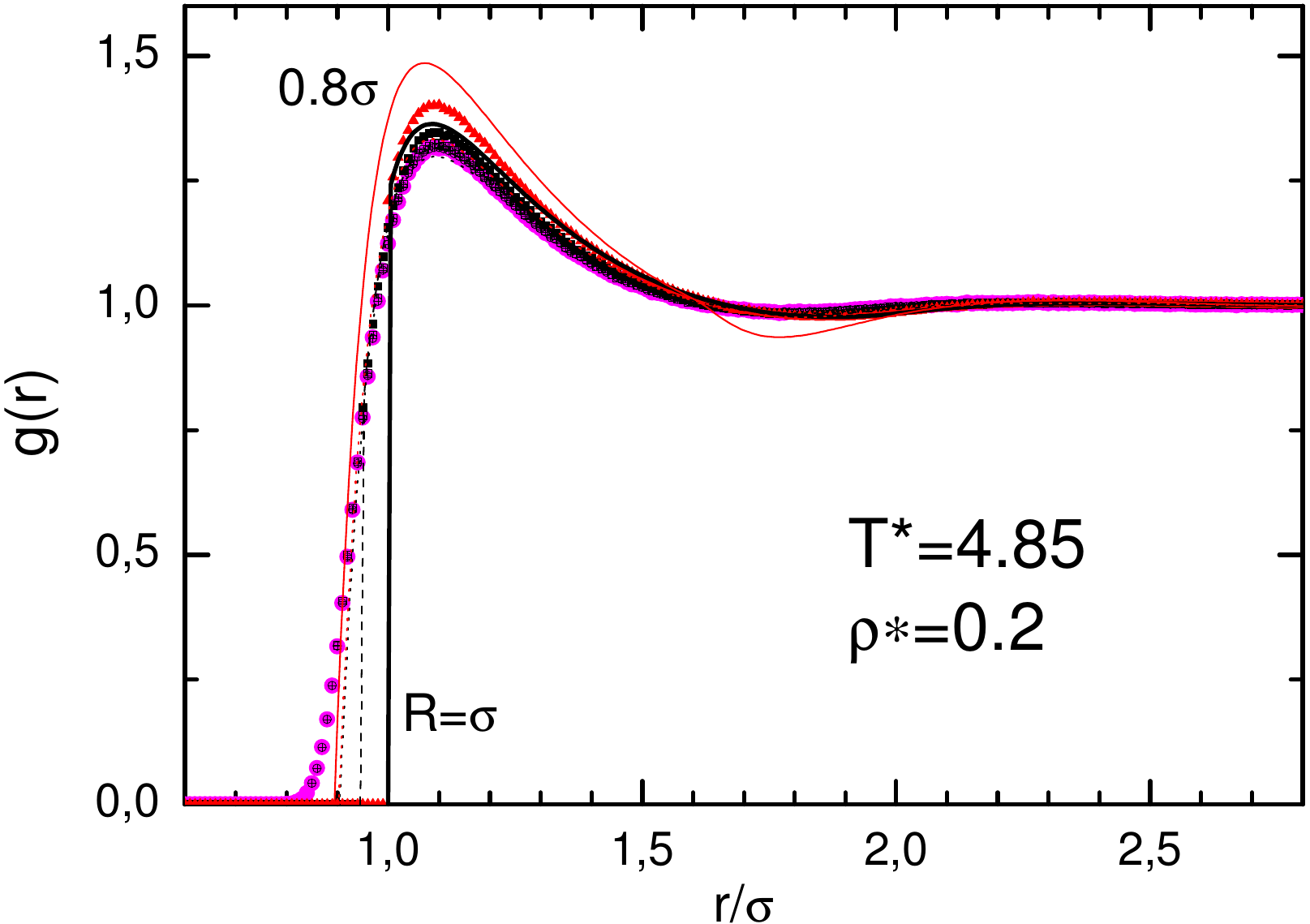}
\includegraphics[width=7cm]{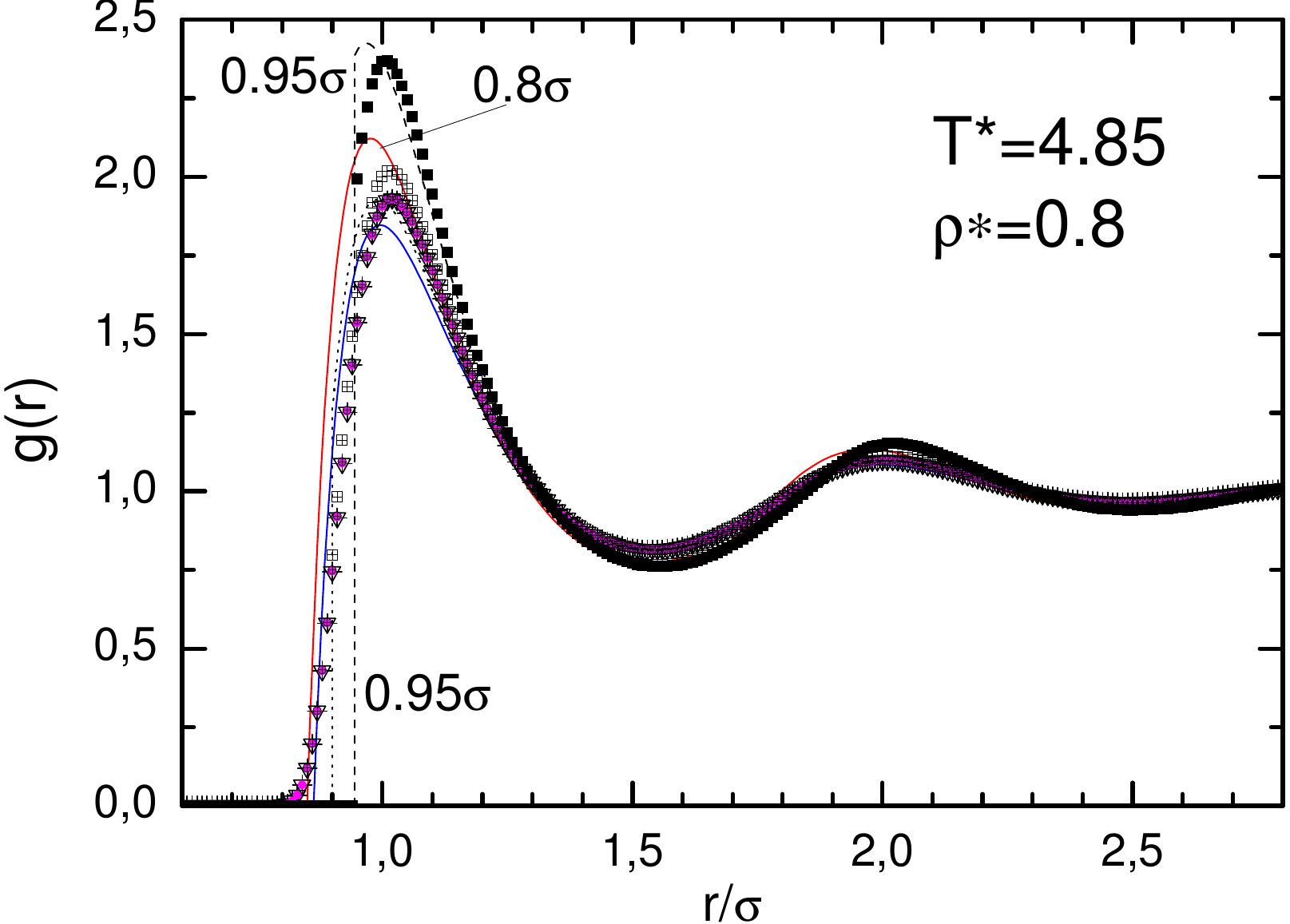}
\end{center}
\caption{The radial distribution functions of the LJ2Y fluid obtained from Monte Carlo computer simulations
(symbols) and  in compliance with the MSA theory (lines) at high temperature $T^*=4.85$ and two densities specified in the figure.
The meaning of the symbols is the same as in figure~\ref{fig3}.}
\label{fig6}
\end{figure}

The studies that we are reporting here have been performed for three reduced temperatures,
 $T^*=1.25, 1.45$ and 4.85. These three temperatures reflect three important
temperature conditions of the parent LJ fluid, namely,
subcritical, supercritical and extreme.
\begin{table}[h]
{\small
\begin{center}
\caption{Monte Carlo data for thermodynamics of the Lennard-Jones-like 2Y fluid (LJ2Y) without any hard-core.}
\vspace{2ex}
\begin{tabular}{rrr r r r}
\hline
\multicolumn{1}{c}{$\sigma_{\mathrm{HS}}$}&\multicolumn{1}{c}{$T^*$}&\multicolumn{1}{c}{$\rho^*$}&\multicolumn{1}{c}{$U^*$}&\multicolumn{1}{c}{$P^*$}&\multicolumn{1}{c}{$Z$}\\
\hline
 0.00 &  0.81 & 0.8645 &    --5.5392 $\pm$     0.0012 &     1.5908 $\pm$     0.0009 &     2.2718 $\pm$     0.0009 \\
 0.00 &  1.25 & 0.70 &    --4.3084 $\pm$     0.0014 &     0.9776 $\pm$     0.0004 &     1.1638 $\pm$     0.0004 \\
 0.00 &  1.25 & 0.80 &    --4.8443 $\pm$     0.0015 &     2.3691 $\pm$     0.0007 &     2.4678 $\pm$     0.0007 \\
 0.00 &  1.25 & 0.85 &    --5.0608 $\pm$     0.0017 &     3.5213 $\pm$     0.0008 &     3.4523 $\pm$     0.0008 \\
 0.00 &  1.25 & 0.90 &    --5.2355 $\pm$     0.0018 &     5.0848 $\pm$     0.0012 &     4.7081 $\pm$     0.0012 \\
 0.00 &  1.45 & 0.30 &    --1.8953 $\pm$     0.0013 &     0.2560 $\pm$     0.0001 &     0.5885 $\pm$     0.0001 \\
 0.00 &  1.45 & 0.40 &    --2.4559 $\pm$     0.0013 &     0.3389 $\pm$     0.0001 &     0.5842 $\pm$     0.0001 \\
 0.00 &  1.45 & 0.50 &    --3.0183 $\pm$     0.0013 &     0.5048 $\pm$     0.0002 &     0.6963 $\pm$     0.0002 \\
 0.00 &  1.45 & 0.70 &    --4.1552 $\pm$     0.0015 &     1.7492 $\pm$     0.0005 &     1.7233 $\pm$     0.0005 \\
 0.00 &  1.45 & 0.95 &    --5.0474 $\pm$     0.0023 &     8.8882 $\pm$     0.0016 &     6.4524 $\pm$     0.0016 \\
 0.00 &  4.85 & 0.20 &    --0.8883 $\pm$     0.0018 &     1.1520 $\pm$     0.0001 &     1.1876 $\pm$     0.0001 \\
 0.00 &  4.85 & 0.40 &    --1.7065 $\pm$     0.0029 &     3.0689 $\pm$     0.0003 &     1.5819 $\pm$     0.0003 \\
 0.00 &  4.85 & 0.60 &    --2.3346 $\pm$     0.0038 &     6.8935 $\pm$     0.0006 &     2.3689 $\pm$     0.0006 \\
 0.00 &  4.85 & 0.80 &    --2.4590 $\pm$     0.0051 &    14.9992 $\pm$     0.0012 &     3.8658 $\pm$     0.0012 \\
 0.00 &  4.85 & 1.00 &    --1.5713 $\pm$     0.0062 &    31.7007 $\pm$     0.0025 &     6.5362 $\pm$     0.0025 \\
\hline
\end{tabular}
\end{center}}
\label{tab2}
\vspace{-2ex}
\end{table}
\begin{table}[!h]
\vspace{-2ex}
{\small
\begin{center}
\caption{Monte Carlo data and MSA results (in parenthesis) for thermodynamics of the Lennard-Jones-like 2Y fluid
(LJ2Y) with a hard-core located at $R=0.8\sigma$.}
\vspace{2ex}
\begin{tabular}{rr r r r}
\hline
\multicolumn{1}{c}{$T^*$}&\multicolumn{1}{c}{$\rho^*$}&\multicolumn{1}{c}{$U^*$}&\multicolumn{1}{c}{$P^*$}&\multicolumn{1}{c}{$Z$}\\
\hline
 1.25 & 0.70 &    --4.2994 $\pm$     0.0419 (--16.3395) &     0.9725 $\pm$     0.0031 (3.2445) &     1.1578 $\pm$     0.0031 (3.7080) \\
 1.25 & 0.80 &    --4.7962 $\pm$     0.0239 (--15.9032) &     2.3756 $\pm$     0.0333 (6.6936) &     2.4746 $\pm$     0.0333 (6.6936) \\
 1.25 & 0.85 &    --5.0644 $\pm$     0.0280 (--15.4926) &     3.4930 $\pm$     0.0042 (8.9445) &     3.4245 $\pm$     0.0042 (8.4181) \\
 1.25 & 0.90 &    --5.2403 $\pm$     0.0051 (--14.9595) &     5.0984 $\pm$     0.0180 (11.5884) &     4.7207 $\pm$     0.0180 (10.3008) \\
 1.45 & 0.30 &    --1.9247 $\pm$     0.0328 (--26.8528) &     0.2564 $\pm$     0.0004 (--2.0151) &     0.5895 $\pm$     0.0004 (--4.6421) \\
 1.45 & 0.40 &    --2.4288 $\pm$     0.0316 (--31.5121) &     0.3387 $\pm$     0.0003 (--2.2172) &     0.5839 $\pm$     0.0003 (--3.8314) \\
 1.45 & 0.50 &    --3.0109 $\pm$     0.0123 (--34.5365) &     0.5109 $\pm$     0.0044 (--1.3935)&     0.7046 $\pm$     0.0044 (--1.9267) \\
 1.45 & 0.70 &    --4.1595 $\pm$     0.0107 (--36.5587) &     1.7429 $\pm$     0.0049 (--4.9188) &     1.7171 $\pm$     0.0049 (--4.8570) \\
 1.45 & 0.95 &    --5.0264 $\pm$     0.0324 (--32.8354) &     8.8919 $\pm$     0.0253 (--26.0895) &     6.4551 $\pm$     0.0253 (--18.9368) \\
 4.85 & 0.20 &    --0.8916 $\pm$     0.0176 (--5.8499) &     1.1504 $\pm$     0.0012 (0.8452) &     1.1860 $\pm$     0.0012 (0.8729) \\
 4.85 & 0.40 &    --1.8134 $\pm$     0.0401 (--9.4717) &     3.0687 $\pm$     0.0006 (2.7345) &     1.5818 $\pm$     0.0006 (1.4126) \\
 4.85 & 0.60 &    --2.2961 $\pm$     0.0130 (--10.6515) &     6.8818 $\pm$     0.0021 (8.0328) &     2.3649 $\pm$     0.0021 (2.7667) \\
 4.85 & 0.80 &    --2.6086 $\pm$     0.0897 (--9.3413) &    15.0001 $\pm$     0.0335 (19.6659) &     3.8660 $\pm$     0.0335 (5.0803) \\
 4.85 & 1.00 &    --1.6613 $\pm$     0.1132 (--5.4406) &    31.7864 $\pm$     0.0342 (41.6984) &     6.5539 $\pm$     0.0342 (8.5976) \\
\hline
\end{tabular}
\end{center}}
\label{tab3}
\end{table}
This can be illustrated by the phase diagram
shown in figure~\ref{fig2} with a reminder that critical point temperature of the LJ fluid
is around $T^*_{\rm c,LJ}\approx 1.35$~\cite{lotfi}.
The distances that have been explored for the position of hard-core diameter are: $R=\sigma$, $0.95\sigma$, $0.9\sigma$, $0.85\sigma$
and $0.8\sigma$. The values of the internal energy $U^*$, pressure $P^*$ and compressibility
factor $Z=P/\rho kT$ for each
hard-core position and at three temperature conditions are collected in tables from 3 to 7.
Additionally, the density dependencies of the pressure $P^*$ and internal energy $U^*$
are illustrated in figures~\ref{fig3} and~\ref{fig4}, respectively. Typical radial distribution functions are
presented in figures~\ref{fig5} and~\ref{fig6}. Tables and figures show both computer simulation data
and  the results obtained from the MSA theory. For the sake of comparison between different parts, figures~\ref{fig3} and~\ref{fig4}
are using the same scale which is  not always convenient for distinguishing the data within the same part of the figure.
For these purposes we recommend to look for the data in tables~2--7.

\looseness=1For the lowest temperature, $T^*=1.25$, which is slightly below the critical temperature  for the
LJ fluid one can see (upper parts in figures~\ref{fig3} and~\ref{fig4}) that MC simulation data indicate very weak dependence of
the thermodynamic data on the replacement of soft repulsion by stiff
hard-core repulsion in the entire range of distances $r\leqslant \sigma$. This is more evident
for pressure, when only the limiting case $R=\sigma$ (filled triangles in figures from~\ref{fig3} to~\ref{fig6})
indicates the tendency to be separated from the rest of data;
in the case of internal energy, similar tendency is seen for the case $R=0.95\sigma$.
Quite similar dependencies on the hard-core position are shown by MC computer simulation data for the higher
temperature, $T^*=1.45$ (see the middle parts in figures~\ref{fig3} and~\ref{fig4}). Indeed, we can see that the case
$R=\sigma$ is rather special for these temperature conditions. Interestingly, this is also true
for the MSA theory (thick solid lines in figures from~\ref{fig3} to~\ref{fig6}). We also note that being compared with MC data
for the case of a hard-core placed at zero-potential energy separation distance $R=\sigma$,
the MSA theory performs quite well for the  thermodynamics of the system. The MSA curves calculated with the hard-core placed at shorter distances, $R=0.95\sigma$, for both pressure and energy
(dashed lines in figures~\ref{fig3} and~\ref{fig4}) are clearly separated from the MSA results obtained with the hard-core at
$R=\sigma$ and are very close to the MC computer simulation data.
\begin{table}[h]
\vspace{-1ex}
{\small
\begin{center}
\caption{The same as in table 3 but with a hard-core located at $R=0.85\sigma$.}
\vspace{0.25cm}
\begin{tabular}{rr r r r}
\hline
\multicolumn{1}{c}{$T^*$}&\multicolumn{1}{c}{$\rho^*$}&\multicolumn{1}{c}{$U^*$}&\multicolumn{1}{c}{$P^*$}&\multicolumn{1}{c}{$Z$}\\
\hline
1.25 & 0.70 &    --4.3081 $\pm$     0.0226 (--7.9827) &     0.9914 $\pm$     0.0081 (2.4220) &     1.1802 $\pm$     0.0081 (2.7680) \\
1.25 & 0.80 &    --4.7836 $\pm$     0.0109 (--7.8873) &     2.3767 $\pm$     0.0104 (4.8594) &     2.4757 $\pm$     0.0104 (4.8594) \\
1.25 & 0.85 &    --5.1362 $\pm$     0.0261 (--7.7041) &     3.5104 $\pm$     0.0043 (6.5203) &     3.4415 $\pm$     0.0043 (6.1366) \\
1.25 & 0.90 &    --5.1569 $\pm$     0.0355 (--7.4408) &     5.1130 $\pm$     0.0144 (8.5280) &     4.7342 $\pm$     0.0144 (7.5805) \\
1.45 & 0.30 &    --1.8984 $\pm$     0.0207 (--6.7843) &     0.2577 $\pm$     0.0015 (--0.1523) &     0.5924 $\pm$     0.0015 (--0.3509) \\
1.45 & 0.40 &    --2.4270 $\pm$     0.0301 (--8.3220) &     0.3377 $\pm$     0.0016 (--0.0368) &     0.5822 $\pm$     0.0016 (--0.0637) \\
1.45 & 0.50 &    --3.0328 $\pm$     0.0127 (--9.4128) &     0.5029 $\pm$     0.0012 (0.5125) &     0.6936 $\pm$     0.0012 (0.7083) \\
1.45 & 0.70 &    --4.1860 $\pm$     0.0193 (--10.1948) &     1.7520 $\pm$     0.0059 (3.9614) &     1.7262 $\pm$     0.0059 (3.9117) \\
1.45 & 0.95 &    --5.0685 $\pm$     0.0069 (--8.4980) &     8.8548 $\pm$     0.0213 (16.2394) &     6.4282 $\pm$     0.0213 (11.7872) \\
4.85 & 0.20 &    --0.9473 $\pm$     0.0343 (--1.5582) &     1.1596 $\pm$     0.0115 (1.1249) &     1.1955 $\pm$     0.0115 (1.1618) \\
4.85 & 0.40 &    --1.7224 $\pm$     0.0409 (--2.6245) &     3.0919 $\pm$     0.0218 (3.1618) &     1.5937 $\pm$     0.0218 (1.6333) \\
4.85 & 0.60 &    --2.4364 $\pm$     0.0392 (--2.9535) &     6.9479 $\pm$     0.0715 (7.5399) &     2.3876 $\pm$     0.0715 (2.5969) \\
4.85 & 0.80 &    --2.6936 $\pm$     0.0909 (--2.2996) &    15.1613 $\pm$     0.0870 (16.6089) &     3.9076 $\pm$     0.0870 (4.2906) \\
4.85 & 1.00 &    --1.7806 $\pm$     0.1141 (--0.3807) &    31.7812 $\pm$     0.4630 (34.3812) &     6.5528 $\pm$     0.4630 (7.0889) \\
\hline
\end{tabular}
\end{center}}
\label{tab4}
\end{table}
\begin{table}[h]
\vspace{-2ex}
{\small
\begin{center}
\caption{The same as in table~3 but with a hard-core located at $R=0.9\sigma$.}
\vspace{0.25cm}
\begin{tabular}{rrr r r r}
\hline
\multicolumn{1}{c}{$T^*$}&\multicolumn{1}{c}{$\rho^*$}&\multicolumn{1}{c}{$U^*$}&\multicolumn{1}{c}{$P^*$}&\multicolumn{1}{c}{$Z$}\\
\hline
 1.25 & 0.70 &    --4.3027 $\pm$     0.0225 (--4.8682) &     0.9804 $\pm$     0.0068 (1.6581) &     1.1672 $\pm$     0.0068 (1.8706) \\
 1.25 & 0.80 &    --4.8347 $\pm$     0.0078 (--5.0781) &     2.3805 $\pm$     0.0212 (3.4166) &     2.4796 $\pm$     0.0212 (3.4166) \\
 1.25 & 0.85 &    --5.1022 $\pm$     0.0361 (--5.1068) &     3.5282 $\pm$     0.0153 (4.6974) &     3.4590 $\pm$     0.0153 (4.4210) \\
 1.25 & 0.90 &    --5.1948 $\pm$     0.0299 (--5.0802) &     5.1086 $\pm$     0.0262 (6.3142) &     4.7302 $\pm$     0.0262 (5.6126) \\
 1.45 & 0.30 &    --1.8667 $\pm$     0.0218 (--2.3832) &     0.2576 $\pm$     0.0015 (0.2301) &     0.5922 $\pm$     0.0015 (0.5301) \\
 1.45 & 0.40 &    --2.4916 $\pm$     0.0246 (--3.0631) &     0.3329 $\pm$     0.0011 (0.3595) &     0.5740 $\pm$     0.0011 (0.6212) \\
 1.45 & 0.50 &    --3.0062 $\pm$     0.0212 (--3.6535) &     0.5020 $\pm$     0.0108 (0.6795) &     0.6924 $\pm$     0.0108 (0.9393) \\
 1.45 & 0.70 &    --4.1425 $\pm$     0.0353 (--4.4270) &     1.7299 $\pm$     0.0155 (2.6462) &     1.7043 $\pm$     0.0155 (2.6130) \\
 1.45 & 0.95 &    --5.0493 $\pm$     0.0142 (--4.2580) &     9.0221 $\pm$     0.0269 (10.8535) &     6.5496 $\pm$     0.0269 (7.8779) \\
 4.85 & 0.20 &    --1.0006 $\pm$     0.0140 (--0.9999) &     1.1575 $\pm$     0.0174 (1.1613) &     1.1933 $\pm$     0.0174 (1.1994) \\
 4.85 & 0.40 &    --1.8827 $\pm$     0.0278 (--1.8953) &     3.0517 $\pm$     0.0853 (3.1656) &     1.5731 $\pm$     0.0853 (1.6353) \\
 4.85 & 0.60 &    --2.8002 $\pm$     0.0563 (--2.5607) &     6.9707 $\pm$     0.1030 (7.2761) &     2.3954 $\pm$     0.1030 (2.5061) \\
 4.85 & 0.80 &    --3.1223 $\pm$     0.0609 (--2.7985) &    15.1610 $\pm$     0.5335 (16.0293) &     3.9075 $\pm$     0.5335 (4.1409) \\
\hline
\end{tabular}
\end{center}}
\label{tab5}
\end{table}
\begin{table}[!h]
\vspace{-2ex}
{\small
\begin{center}
\caption{The same as in table~3 but with a hard-core located at $R=0.95\sigma$.}
\vspace{0.25cm}
\begin{tabular}{rrr r r r}
\hline
\multicolumn{1}{c}{$T^*$}&\multicolumn{1}{c}{$\rho^*$}&\multicolumn{1}{c}{$U^*$}&\multicolumn{1}{c}{$P^*$}&\multicolumn{1}{c}{$Z$}\\
\hline
 1.25 & 0.70 &    --4.3871 $\pm$     0.0092 (--4.3674) &     0.9627 $\pm$     0.0137 (1.2917) &     1.1461 $\pm$     0.0137 (1.4763) \\
 1.25 & 0.80 &    --4.9808 $\pm$     0.0317 (--4.8345) &     2.4198 $\pm$     0.0559 (2.8369) &     2.5206 $\pm$     0.0559 (2.8369) \\
 1.25 & 0.85 &    --5.1823 $\pm$     0.0216 (--5.0266) &     3.6847 $\pm$     0.0610 (4.0519) &     3.6124 $\pm$     0.0610 (3.8134)\\
 1.45 & 0.30 &    --1.9166 $\pm$     0.0214 (--1.8592) &     0.2739 $\pm$     0.0162 (0.2652) &     0.6297 $\pm$     0.0162 (0.6108) \\
 1.45 & 0.40 &    --2.5016 $\pm$     0.0075 (--2.4347) &     0.3106 $\pm$     0.0138 (0.3645) &     0.5355 $\pm$     0.0138 (0.6299) \\
 1.45 & 0.50 &    --3.1156 $\pm$     0.0266 (--3.0047) &     0.5126 $\pm$     0.0249 (0.5765) &     0.7071 $\pm$     0.0249 (0.7969) \\
 1.45 & 0.70 &    --4.2619 $\pm$     0.0286 (--4.0737) &     1.7885 $\pm$     0.1040 (2.0262) &     1.7621 $\pm$     0.1040 (2.0008) \\
 4.85 & 0.20 &    --1.1121 $\pm$     0.0170 (--1.0698) &     1.1870 $\pm$     0.0396 (1.1943) &     1.2237 $\pm$     0.0396 (1.2335) \\
 4.85 & 0.40 &    --2.2148 $\pm$     0.0142 (--2.1748) &     3.1669 $\pm$     0.0449 (3.3415) &     1.6324 $\pm$     0.0449 (1.7261) \\
 4.85 & 0.60 &    --3.3309 $\pm$     0.0147 (--3.2611) &     7.4178 $\pm$     0.2012 (7.9735) &     2.5491 $\pm$     0.2012 (2.7463) \\
 4.85 & 0.80 &    --4.3372 $\pm$     0.0376 (--4.1799) &    19.1784 $\pm$     0.5030 (18.6856) &     4.9429 $\pm$     0.5030 (4.8271) \\
\hline
\end{tabular}
\end{center}}
\vspace{-3ex}
\label{tab6}
\end{table}

\begin{table}[h]
{\small
\begin{center}
\caption{The same as in table~3 but with a hard-core located at $R=\sigma$.}
\vspace{0.25cm}
\begin{tabular}{rr r r r}
\hline
\multicolumn{1}{c}{$T^*$}&\multicolumn{1}{c}{$\rho^*$}&\multicolumn{1}{c}{$U^*$}&\multicolumn{1}{c}{$P^*$}&\multicolumn{1}{c}{$Z$}\\
\hline
 1.25 & 0.70 &    --4.6285 $\pm$     0.0201 (--4.5755) &     1.4418 $\pm$     0.1111 (1.5986)&     1.7164 $\pm$     0.1111 (1.8270) \\
 1.45 & 0.30 &    --1.9693 $\pm$     0.0031 (--1.8995) &     0.3043 $\pm$     0.0362 (0.2881) &     0.6996 $\pm$     0.0362 (0.6636) \\
 1.45 & 0.40 &    --2.5870 $\pm$     0.0119 (--2.5222) &     0.3813 $\pm$     0.0258 (0.4013) &     0.6575 $\pm$     0.0258 (0.6934) \\
 1.45 & 0.50 &    --3.2332 $\pm$     0.0154 (--3.1744) &     0.5695 $\pm$     0.0126 (0.6383) &     0.7855 $\pm$     0.0126 (0.8824) \\
 1.45 & 0.70 &    --4.6221 $\pm$     0.0079 (--4.5500) &     2.2816 $\pm$     0.0362 (2.3683) &     2.2479 $\pm$     0.0362 (2.3387) \\
 4.85 & 0.20 &    --1.1368 $\pm$     0.0137 (--1.1383) &     1.1796 $\pm$     0.0193 (1.2640) &     1.2161 $\pm$     0.0193 (1.3056) \\
 4.85 & 0.40 &    --2.4041 $\pm$     0.0068 (--2.4050) &     4.1509 $\pm$     0.2956 (3.7915) &     2.1396 $\pm$     0.2956 (1.9586) \\
 4.85 & 0.60 &    --3.7846 $\pm$     0.0061 (--3.7873) &    10.0931 $\pm$     0.4284 (9.8521) &     3.4684 $\pm$     0.4284 (3.3933) \\
\hline
\end{tabular}
\end{center}}\label{tab7}
\end{table}

At the highest temperature, $T^*=4.85$, (see the bottom parts in figures~\ref{fig3} and \ref{fig4}) that we are referring to as
the extremely high for the LJ fluid (more than three
times higher than its critical point temperature), simulation
data  show that fixing a hard-core at distances $r\leqslant 0.9\sigma$  produces, for all the calculated thermodynamic properties, the values that are very close to those
for the initial LJ2Y model without a hard-core. At the same time, hard-core $R=0.9\sigma$ seems to be the
smallest one when the MSA theory performs well. Although this is not the case of the pressure (bottom part in figure~\ref{fig3}),
but this is already seen from the behavior of the internal energy (bottom part in figure~\ref{fig4}) and, especially, from the
comparison with MC data for the radial distribution function (figures~\ref{fig5} and \ref{fig6}) when extra oscillations around
the first minima start to appear.

\section{Conclusions}

In this study we report the Monte Carlo (MC) simulation studies of the thermodynamics and radial distribution
functions of the Lennard-Jones-like two Yukawa (LJ2Y) fluid. This fluid model has been defined to nearly exactly mimic
 the pair interaction at short and intermediate distances in the Lennard-Jones (12--6) fluid.
In particular, the parameters of the LJ2Y have been defined from the condition that zero-potential energy distance and
the slope of the potential profile at the zero-potential distance are the same.  Moreover, the magnitude and position
of the potential minima in both models coincide. The only differences that
two potential models experience are attributed to large distances. Exactly these differences have caused quite
pronounced differences in the vapour-liquid coexistence occurring in two models. However, the purpose of the present
study is not to describe the LJ fluid by means of the two Yukawa model. Our goal is to find how the mean-spherical
approximation (MSA)  performs when it is applied to the fluid model whose soft repulsion is the same or quite similar to
that of the LJ fluid. The task we set forth is twofold. First, the application of the MSA theory requires cutting-off
the soft repulsion at some interparticle separation and placing the hard-core $R$ at this distance.
However, only such a modification of the initial potential may intervene and change the properties of the system.
Without this knowledge it will be hard to judge about the MSA performance. To fulfill this task, we performed
MC computer simulations for both the initial LJ2Y model defined according to equation~(\ref{u2Y}) and for hard-core modified
potential (\ref{u2Ymsa})~-- the form that is usually utilized by the MSA theory. The positions of the hard-core that
have been implemented in computer simulations include $R=\sigma;$ $0.95\sigma;$ $0.9\sigma;$ $0.85\sigma$ and $0.8\sigma$.
Having compared this with the data obtained in computer simulations with initial LJ2Y potential, i.e., without hard-core,
we conclude that the insertion of the hard-core into a LJ-like models is extremely sensitive to the temperature conditions.
Namely, at normal temperature conditions of the order and around the critical point temperature, an insertion of
the hard-core at separations $r\leqslant 0.95\sigma$ practically does not change the thermodynamics of the initial system,
while the case $R=\sigma$ already shows the tendency to exhibit slightly different thermodynamic properties.
The conclusions change when we explored the high temperature conditions, namely, $T^*=4.85$. At these temperature
conditions, the presence of the hard-core at distances as short as $r\leqslant 0.85\sigma$ may be considered  not to modify the initial system. All other cases, $R=\sigma;$ $0.95\sigma$ and $0.9\sigma$ should  be treated as
separate model fluids that are different from the initial LJ2Y fluid model. These observations are valid for both
thermodynamics and radial distribution functions.

Having taken these considerations into account we were able to fulfill the second part of our task, namely, to judge the
MSA performance for the LJ2Y fluid model. First of all, we found that MSA correctly reflects the temperature conditions
when the case of a hard-core $R=\sigma$ starts to deviate from the other hard-core modified models. Secondly, the MSA
performs rather satisfactorily for this case $R=\sigma$ at all considered temperature conditions. The following conclusion
seems to be rather important and reads: at normal temperature conditions (of the order and around
the critical point temperature for the LJ fluid), an optimal position of the hard-core $R$ for the MSA to be a
reasonable theoretical approach is $R=0.95\sigma$. This means that placing the hard-core at $R=0.95\sigma$
and using the MSA theory to describe the  LJ-like two Yukawa model will give you the results that are of reasonable
accuracy and are very close to those of the initial model. By contrast, using the hard-core for MSA description as small as
$R\leqslant 0.9\sigma$ will result in incorrect values for initial LJ2Y fluid model and even for the LJ-like model
with an embedded hard-core.

The above conclusion is valid when you are seeking the possibility to use the MSA theory at high temperatures, $T^*=4.85$.
The only thing that changes in this case is the limiting value of the hard-core. Namely, at this high temperature you may
use the hard-core position as small as $R=0.9\sigma$ and still continue to apply the MSA theory and obtain reasonably
accurate description of the LJ2Y fluid model. On the contrary, using the hard-core at $R=0.85\sigma$ and $0.8\sigma$
leads to an increased inaccuracy in both thermodynamics and the radial distribution functions. We also note that at
this high temperature, the hard-core modified LJ2Y models with $R=\sigma$ and $R=0.95\sigma$ significantly differ
from the initial LJ2Y model without hard-core and, very importantly, the MSA theory performs quite well for both these
cases.


\section*{Acknowledgement}

This work was supported by the Grant Agency of the Academy of Sciences
of the Czech Republic (Grant No. IAA400720710) and the Czech-Ukrainian Bilateral Cooperative Program.

\ukrainianpart

\title{Середньосферичне наближення для Леннард-Джонс-\\подібного плину, змодельованого сумою двох потенціалів Юкави: порівняння з результатами методу Монте Карло}

\author{Я. Крейчi\refaddr{IOP}, I. Незбеда\refaddr{IOP,LCI},  Р. Мельник\refaddr{IP}, А. Трохимчук\refaddr{IP,BYU}}

\addresses{
\addr{IOP} Факультет природничих наук, Унiверситет Я.Е. Пуркiнйє, Устi над Лабем, Чеська Республiка
\addr{LCI} Iнститут фундаментальних основ хiмiчних процесiв, Академiя наук, Прага,
Чеська Республiка
\addr{IP} Інститут фізики конденсованих систем НАН України,
 вул. І.~Свєнціцького, 1, 79011 Львів, Україна
\addr{BYU}  Факультет хімії та біохімії, Університет Бригам Янг, Прово, США
}

\makeukrtitle

\begin{abstract}
\tolerance=3000
Методом Монте Карло проведено дослідження Леннард-Джонс-подібного плину, взаємодія між частинками якого задається сумою двох потенціалів Юкави (ЛД2Ю). Метою досліджень є продемонструвати вплив, якому піддаються властивості моделі при заміні ``м'якого'' відштовхування ``твердим'' кором. Розглянуто різні відстані для розміщення твердого кору. Виявлено, що при температурах, трохи нижчих і трохи вищих за критичну температуру леннард-джонсівського плину, переміщення твердого кору на відстані, коротші за відстань з нульовою потенціальною енергією, практично не змінює термодинамічних властивостей моделі, значення яких є дуже близькими до термодинамічних характеристик ЛД2Ю моделі без твердого кору. Однак, при переході в зону екстремально високих температур слід бути обережним, оскільки присутність твердого кору провокує суттєві зміни властивостей системи. Це застереження є дуже важливим при застосуванні методу середньосферичного наближення до опису Леннард-Джонс-подібного плину.

\keywords  2-Юкава потенціал, леннард-джонсівський плин, середньосферичне наближення, метод  Монте Карло

\end{abstract}

\end{document}